\documentclass[twocolumn,amsmath,amssymb, aps]{revtex4-1}

\pdfoutput=1

\usepackage{graphicx}
\usepackage[usenames]{color}
\usepackage{dcolumn}
\usepackage{bm}
\usepackage{hyperref}
\usepackage[mathlines]{lineno}
\usepackage[caption=false]{subfig}

\hypersetup{
	colorlinks = true,
	urlcolor   = blue,
	linkcolor  = blue,
	citecolor  = blue
}

\begin{document}

\title{Laser-written polarizing directional coupler with reduced interaction length}

\author{I.V.~Dyakonov}
	\email{iv.dyakonov@physics.msu.ru}
\author{M.Yu.~Saygin}
\author{I.V.~Kondratyev}
\author{A.A.~Kalinkin}
\author{S.S.~Straupe}
\author{S.P.~Kulik}

\affiliation{Faculty of Physics, M.V.~Lomonosov Moscow State University, Leninskije Gory 1, Moscow, Russia, 119991}

\date{\today}


\begin{abstract}
	Integrated optical devices are becoming a common tool in modern optical science and engineering. This devices should be designed to allow for precise control and manipulation of light. The femtosecond laser written (FSLW) photonic chips have proven to cope with the most current demands of integrated photonics. However, up to date the polarization degree of freedom has rarely been exploited in experiments performed on the FSLW integrated platform. The main obstacle is an intrinsically low anisotropy of laser written waveguides, preventing the design of polarizing integrated devices with small footprint. In this Letter we demonstrate the approach based on stress-induced anisotropy allowing us to decrease the size of polarizing directional couplers fabricated with the FSLW technology by almost an order of magnitude. We provide an accurate account for the effects emerging in waveguides written in a close proximity to each over.
\end{abstract}

\maketitle

A constantly increasing complexity of modern optical experiments necessitates the transition to an integrated architecture meeting the stringent requirements of scalability and robustness, which cannot be fulfilled by bulk-optics setups \cite{Ralph2017}. Integrated optical technology has found numerous successful applications in classical \cite{siliconPhotonics} and quantum optics \cite{OBrien2013}, and optical communications \cite{Doerr2015}. Among various technologies used for fabrication of integrated optical devices femtosecond laser writing (FSLW) is especially suitable for rapid prototyping using the facilities commonly available in an optical laboratory \cite{DellaValle2008}. For this reason the FSLW technique is popular in quantum optics community, where it was successfully used to produce complex circuits required for experiments in quantum information processing \cite{Marshall2009,Szameit2015} and computing \cite{Crespi2013}.

An ability to realize polarization transformations is desirable in any integrated architecture. This is especially valuable for quantum applications, where control over polarization may pave the way for performing operations with polarization-encoded qubits and, for example, allow for on-chip experiments with hyper-entangled states \cite{MataloniLSA2016}. Some source of birefringence is required to enable control over polarization in FSLW waveguides. This birefringence may come from several sources: it may be determined by the anisotropic shape of the waveguides and the guided modes \cite{KapronJQE1972}, may arise from the specific nano-grating structures induced by laser machining \cite{KazanskyOL2004}, or be induced by the anisotropic mechanical stress of the material in the process of fabrication \cite{McMillenOE2015}. Typically, all sources are present and birefringence of the waveguiding structure results from the interplay of these contributions. 

Integrated devices for polarization control based on the FSLW technology were reported by several groups. These include polarizing directional couplers (analogous to polarizing beam-splitters in bulk optics) \cite{FernandesOE2011,CrespiArXiv2017} and partially polarizing couplers \cite{CrespiNatureComm2011}, as well as birefringent retarders and waveplates \cite{FernandesOE2011waveplate,SzameitSciRep2014}. Unfortunately, achievable birefringence in FSLW waveguides is very low, so the polarization-sensitive components have to be large, which significantly hinders their use in complex integrated circuits. Several attempts to enhance the anisotropic behaviour have been reported \cite{FernandesOE2012stress,SzameitSciRep2014}, but still polarizing integrated devices occupy sufficiently larger space on a chip than the circuits employed to manipulate the photon path degree of freedom. Moreover, increasing the intrinsic birefringence of the waveguide may not be the best solution, since that inevitably leads to undesirable polarization rotation in the connecting segments and straight waveguides. Ideally one wants to localize the area of high birefringence to the active area of the device, while keeping the birefringence of the waveguides low.

In this Letter we demonstrate polarizing directional couplers (PDCs) in fused silica with very strong dependence of the coupling coefficient on polarization. We exploit the anisotropic mechanical stress induced by FSLW by reducing the distance between the coupled waveguides as much as possible. We achieve the length of the completely polarizing coupler of $3.7$~mm, which is almost an order of magnitude smaller than previously reported. Importantly, the waveguides themselves have low birefringence outside the strongly coupled region. We study the evolution of the coupled waveguide modes in the presence of anisotropic and irregular stress in closely spaced waveguides and provide a model which may explain the observed phenomena.

The transmission properties of the PDCs are tuned by the proper choice of the interaction length $L$ and the distance $d$ between the adjacent waveguides inside the interaction region. In analogy with a bulk-optical beamsplitter we define the effective transmission $T$ and reflection $R$ coefficients for an integrated PDC as the ratio of the power at the corresponding output ports to the input power. The dependency of $T$($R$) on the interaction length yields, according to the coupled mode theory \cite{Snyder_Love}, the coupling coefficients $C_{H}$ and  $C_{V}$ inside the interaction region between the closely spaced waveguide eigenmodes. Anisotropic guiding properties give rise to the difference between the coupling coefficients $C_{H}$ and $C_{V}$ due to deviations in both the effective mode index and the spatial mode profile:
\begin{equation}
\label{eq:coupling_C}
		C_{H,V} \sim 
		\iint_{\mathbb{R}^{2}} \Delta n_{H,V}^{2}\left(x,y\right) u_{1H,V}\left(x,y\right)u_{2H,V}\left(x,y\right)\,dx \,dy
\end{equation}
where $\Delta n_{H,V}\left(x,y\right)$ is the refractive index contrast of $H$($V$) polarized mode respectively and $u_{jH,V}\left(x,y\right)$ are the corresponding spatial mode functions. As a consequence the beat lengths of $H$ and $V$ modes differ, allowing to choose an optimal interaction length of the PDC corresponding to the distribution of orthogonally polarized states of light to distinct output modes of the coupler. FSLW directional couplers are typically designed with weak coupling, thus enlarging the beat length and also exhibit rather low induced anisotropy on the order of $\Delta\,n_{anis} \sim 10^{-6}$ \cite{SzameitSciRep2014,SansoniPRL2010lowbiref}. Recent works \cite{SzameitSciRep2014,FernandesOE2012stress} have demonstrated the potential of fused silica for an order of magnitude higher anisotropic structures fabrication. Their approach is based on stress-induced anisotropy emerging in the waveguiding structures in the presence of a non-guiding defect close to the waveguide core. We instead intend to demonstrate strong anisotropic coupling in a system of two closely spaced optical waveguides. We seek for an optimal geometry of a PDC to shorten the interaction length required to spatially separate the orthogonal polarization states.
The interaction length corresponding to a perfect splitting of orthogonal polarization states is
\begin{equation}
\label{eq:L_pdc}
	L_{pdc} = \frac{\pi}{2\left(C_{V}-C_{H}\right)},
\end{equation} 
which gives an exact value of $L_{pdc}$. In order to minimize the footprint of the PDC structure the interacting optical modes have to be brought as close as possible, thus increasing the absolute values of $C_{H}$ and $C_{V}$ to resolve the anisotropic behaviour on a smaller scale. Moreover, shrinking the distance between the waveguides at some point brings up stress-induced anisotropic effects described in \cite{SzameitSciRep2014,FernandesOE2012stress}. The mechanical stress field present in the material after the inscription of the first waveguide of a directional coupler affects the writing conditions for a second waveguide, resulting in increased anisotropy and propagating constants mismatch between the modes guided in two adjacent structures.

\begin{figure}[htbp]
	\subfloat[]{
		\includegraphics[width=0.225\linewidth]{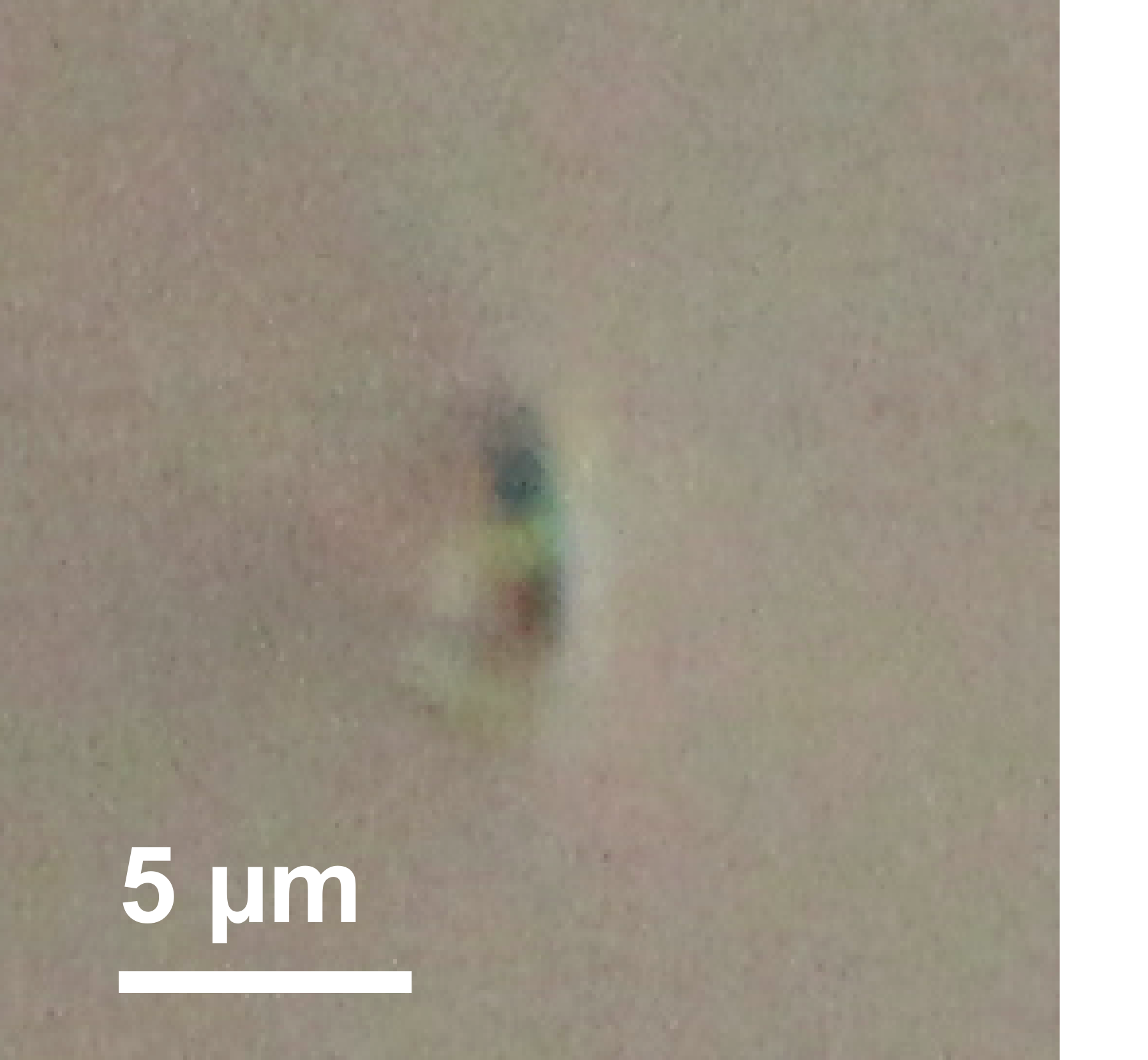}
		\label{fig:cross_section}
		}
	\subfloat[]{
		\includegraphics[width=0.225\linewidth]{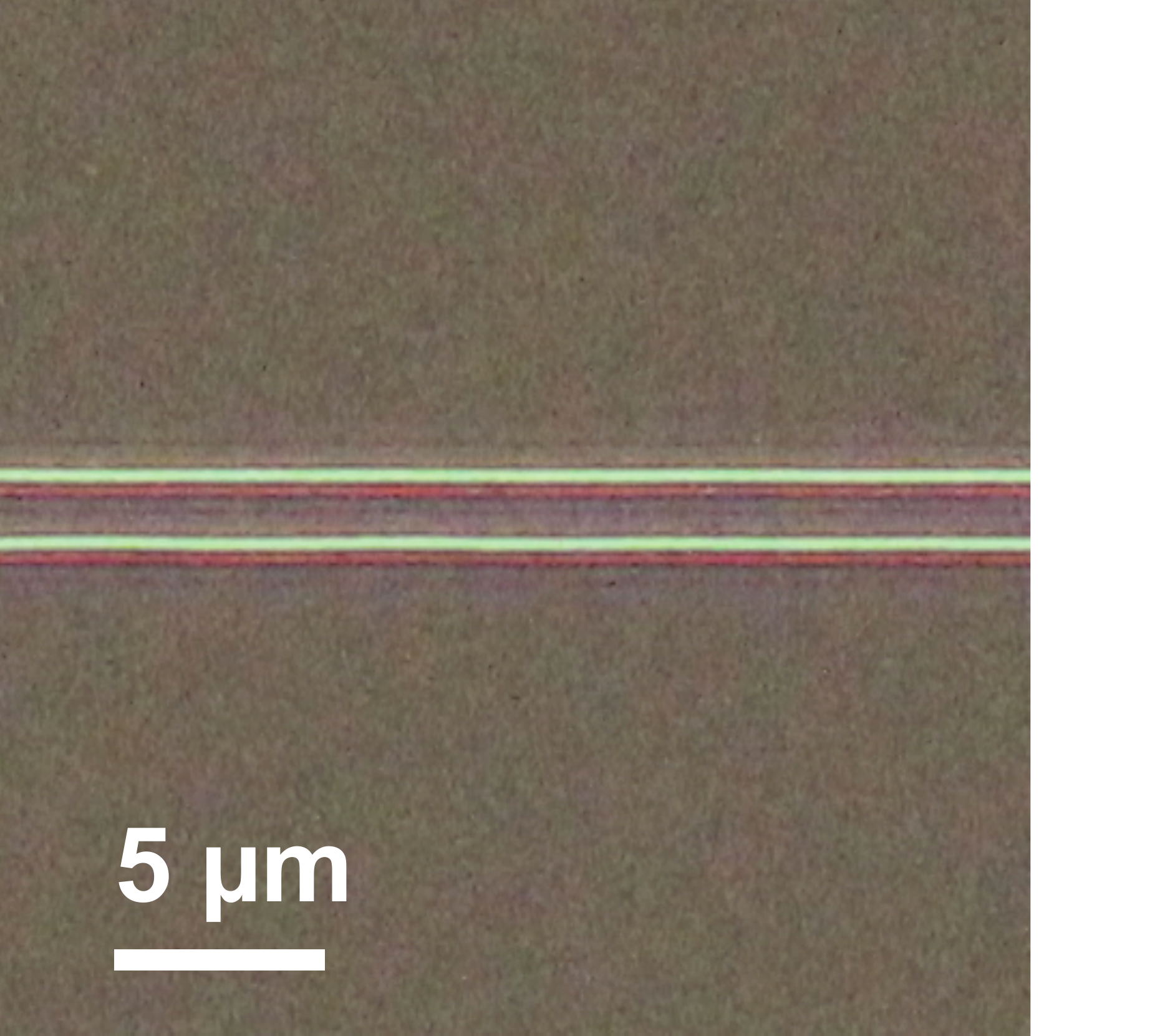}
		\label{fig:interaction}
	}
	\subfloat[]{
		\includegraphics[width=0.225\linewidth]{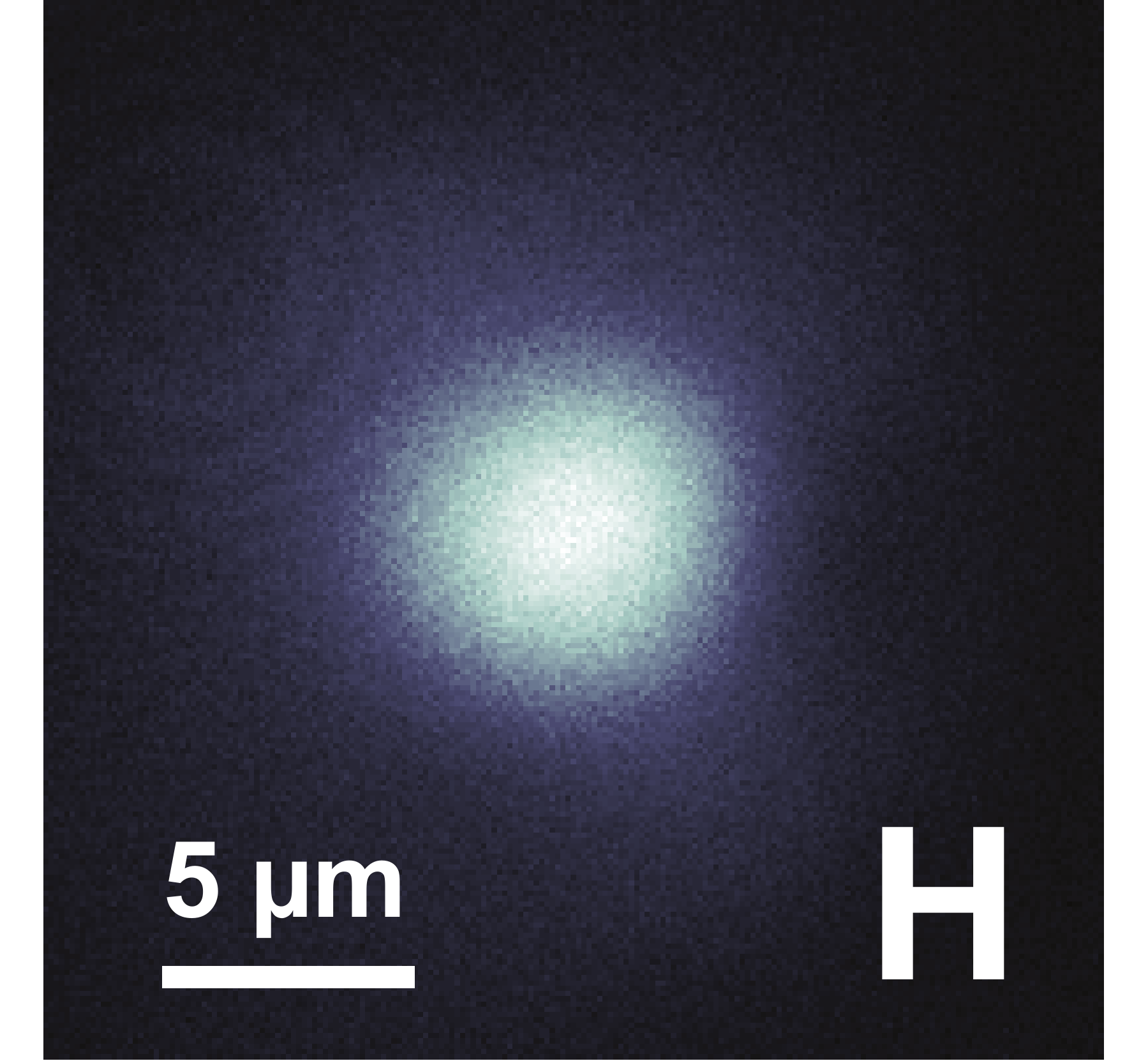}
		\label{fig:h_mode}
	}
	\subfloat[]{
		\includegraphics[width=0.225\linewidth]{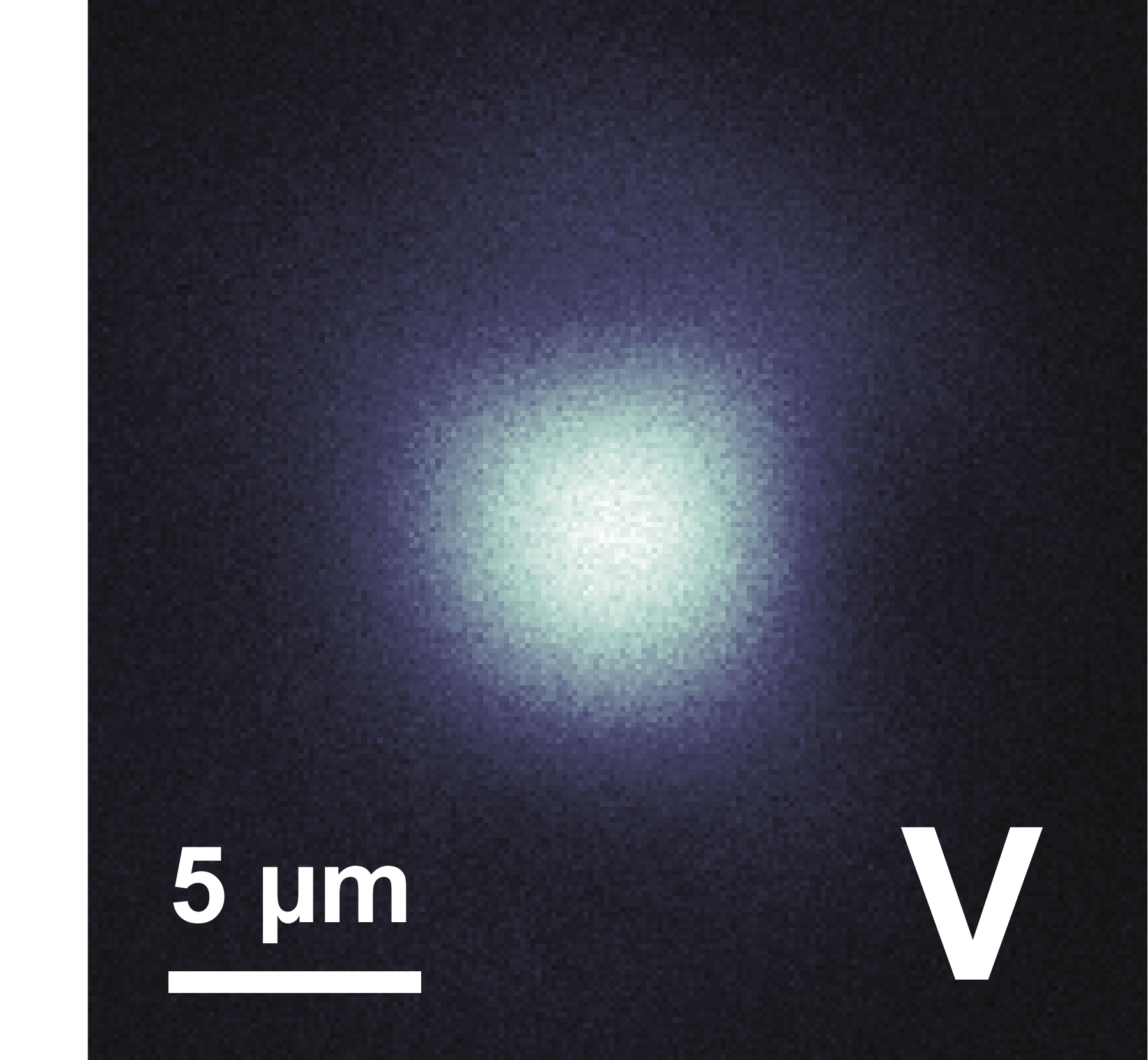}
		\label{fig:v_mode}
	}	
\caption{DIC microscope images of the fabricated waveguide cross-section  (\ref{fig:cross_section}) and the interaction region of the $d = 3$ $\mu$m directional coupler (\ref{fig:interaction}). Images of the horizontal (\ref{fig:h_mode}) and vertical (\ref{fig:v_mode}) waveguide eigenmode profiles, respectively. The analysis of the mode profiles yielded that the fabricated waveguides exhibit almost vanishing intrinsic birefringence on the order of $10^{-6}$.}
\label{fig:sections_and_profiles}
\end{figure}

For our purposes we exploited 50$\times$50$\times$5 mm fused silica slabs (JGS1 glass, AG Optics). 
Such thick glass slabs are mandatory to guarantee that the sample surface flatness variations don't affect the focusing conditions and allows writing the waveguides at small depths of $\sim$20~$\mu$m. 
The sample is exposed to tightly (0.7 NA) focused 400 fs pulses from the frequency doubled ytterbium fiber laser (515 nm) with $\approx 80$ nJ pulse energy delivered at 3 MHz repetition rate. To achieve uniform modification the polarization of the impinging writing beam is oriented parallel to the waveguide writing direction, and the sample is translated along the focal spot at a moderate feed rate of 0.5 mm/s. The details of our fabrication setup may be found in \cite{Dyakonov2016}. The birefringence axis of the waveguide is oriented perpendicular to the sample surface. The waveguide cross-section and mode field profiles for the two orthogonal polarizations are shown in Fig.\ref{fig:sections_and_profiles}.

\begin{figure}[htb]
	\subfloat[$d=3$ $\mu$m]{
		\includegraphics[width=0.49\linewidth]{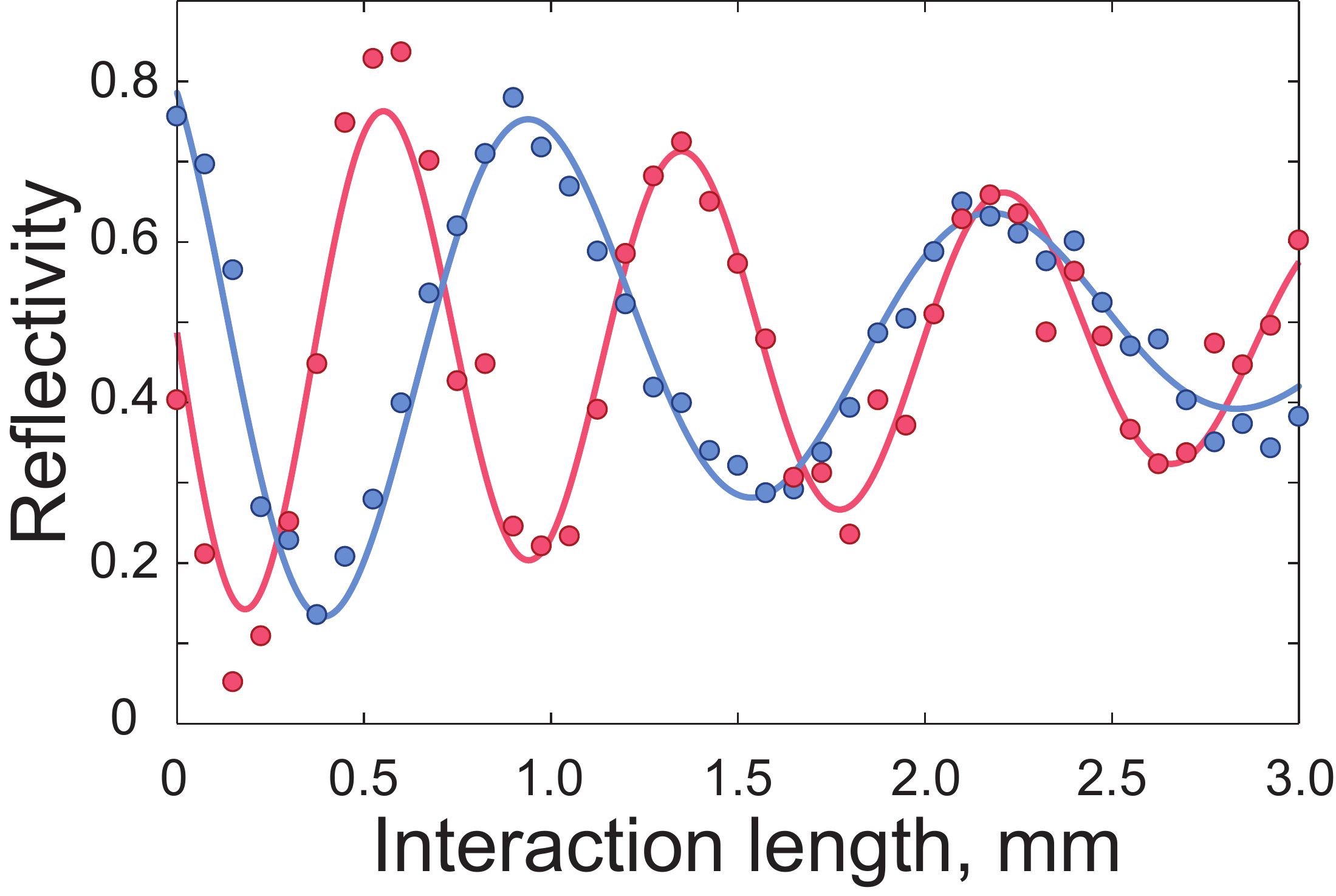}
		\label{fig:3mkm_pdc}
	}
	\subfloat[$d=5$ $\mu$m]{
		\includegraphics[width=0.49\linewidth]{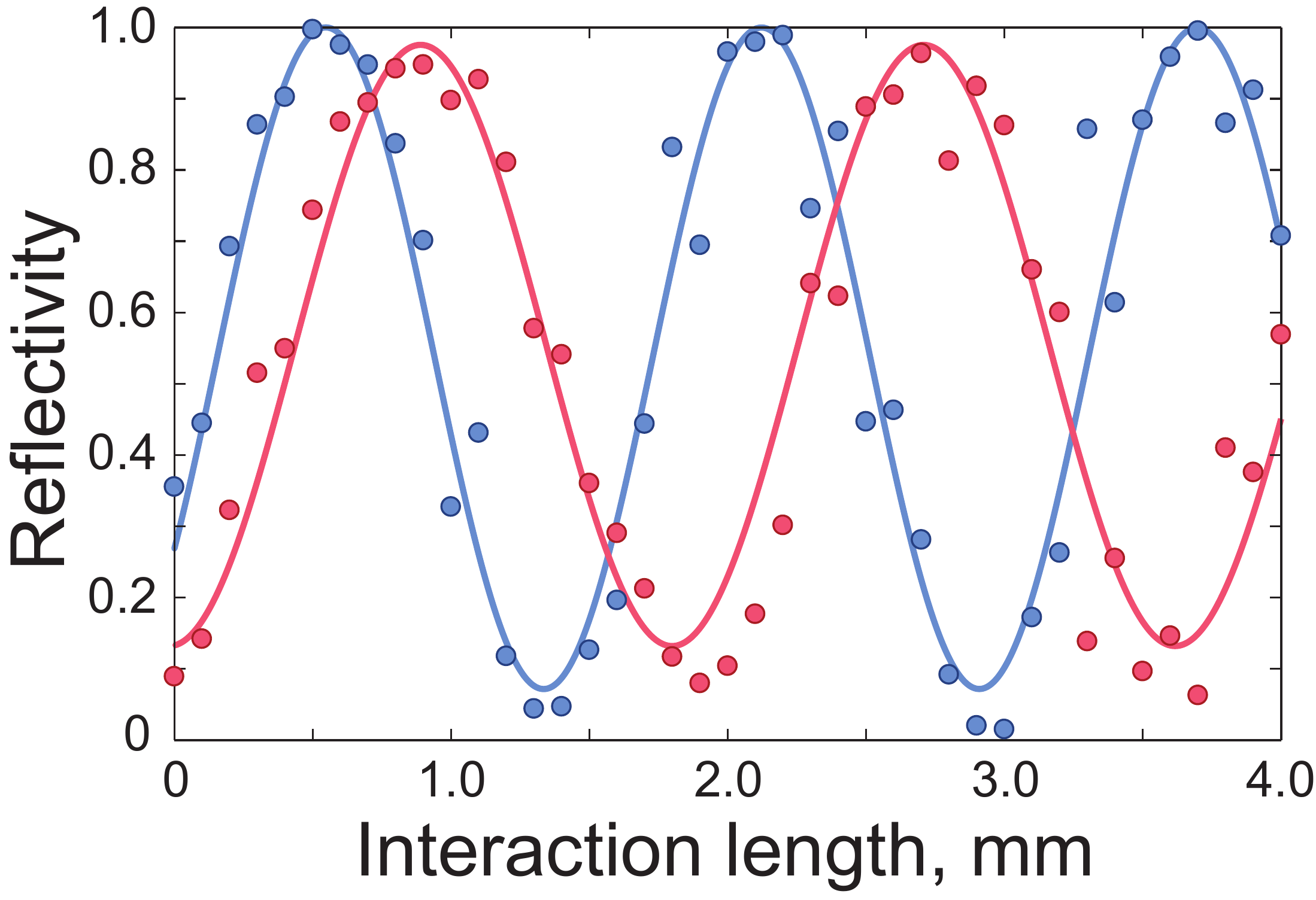}
		\label{fig:5mkm_pdc}
	}\\
	\subfloat[$d=7$ $\mu$m]{
		\includegraphics[width=0.49\linewidth]{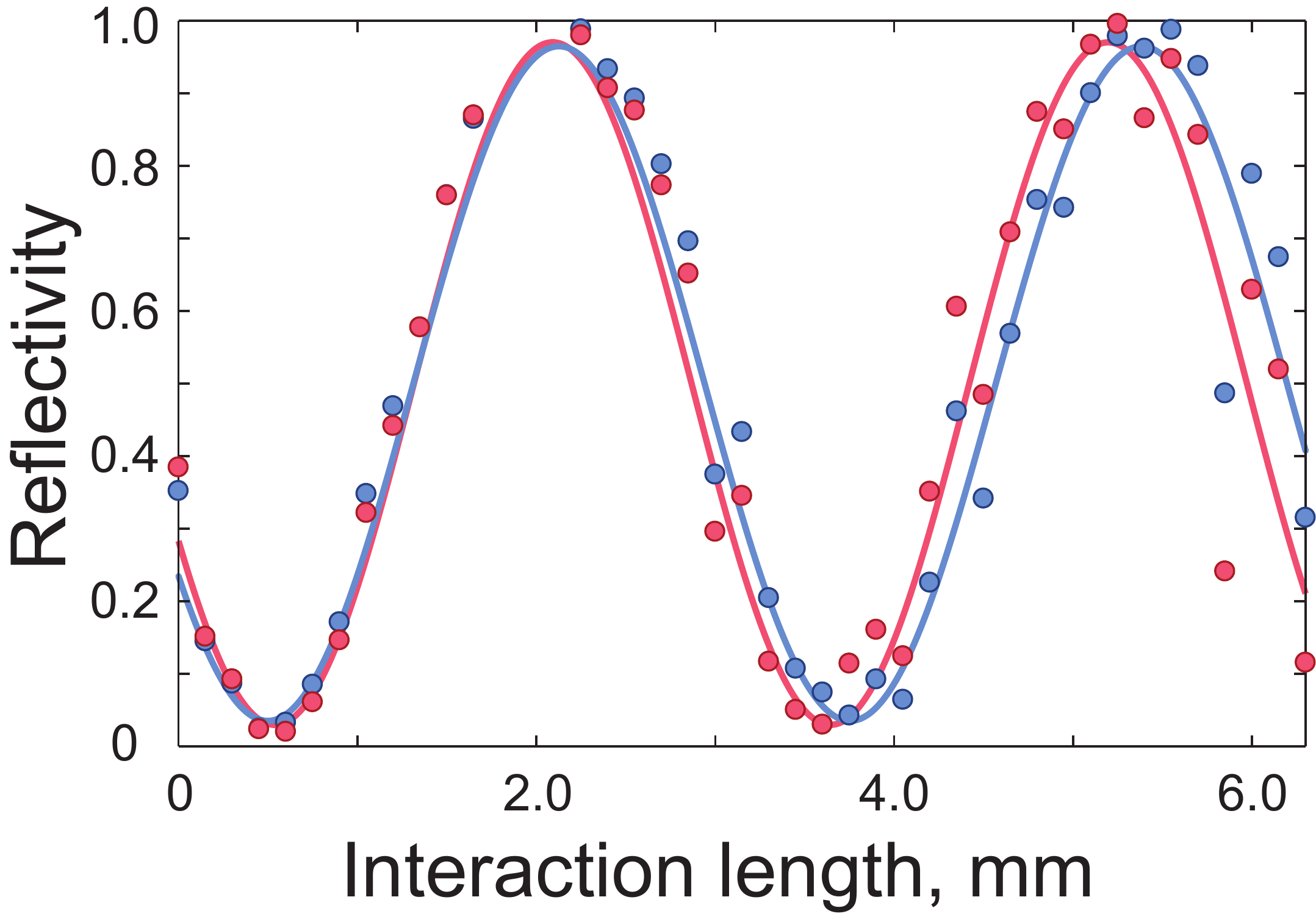}
		\label{fig:7mkm_pdc}
	}
	\subfloat[]{
		\includegraphics[width=0.49\linewidth]{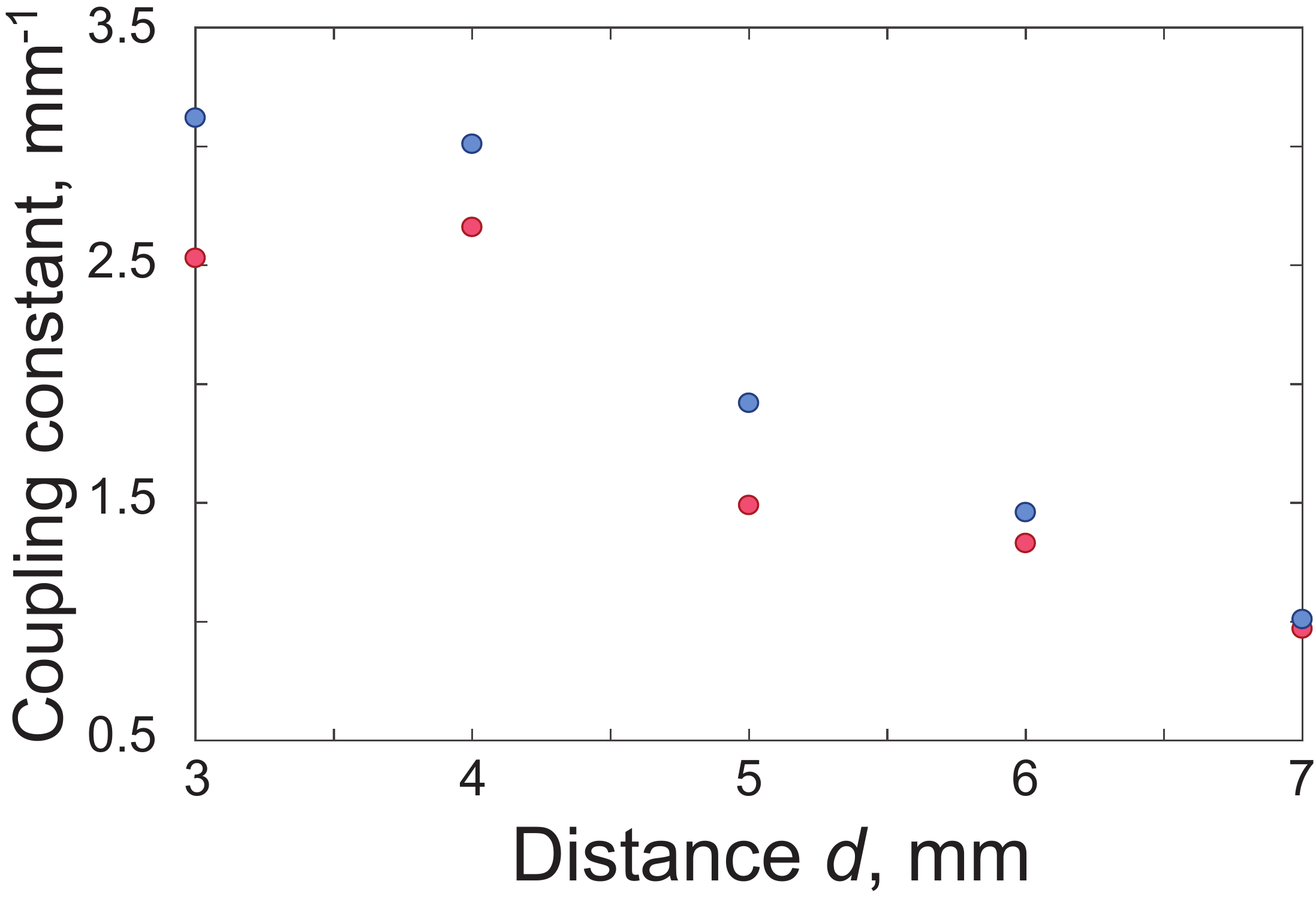}
		\label{fig:coupling_vs_distance}
	}
	\caption{Measured reflection coefficients for the arrays of the directional couplers fabricated with different distance $d$ in the interaction region. Figs.~\ref{fig:3mkm_pdc}--\ref{fig:7mkm_pdc} illustrate the anisotropic coupling inside the interaction region (horizontal polarization -- red color, vertical -- blue). Strong anisotropy is clearly seen at \ref{fig:3mkm_pdc},\ref{fig:5mkm_pdc}, corresponding to closely spaced waveguides, whereas Fig.~\ref{fig:7mkm_pdc} shows weak anisotropic coupling due to the lack of stress-induced effects. Dependence of the coupling coefficients $C_{H,V}$ on the distance between the waveguides inferred from the theoretical fit is shown in Fig.~\ref{fig:coupling_vs_distance}. For the $d = 3$~$\mu$m directional coupler the fit returns $D_{H} = 0.6$~mm$^{-1}$, $D_{V} = 0.383$~mm$^{-1}$ and $\beta_{H} = 4.63$~mm$^{-1}$, $\beta_{V} = 6.56$~mm$^{-1}$; for the $d = 5$~$\mu$m structures $D_{H,V}$ vanish and $\beta_{H} = 1.74$~mm$^{-1}$, $\beta_{V} = 1.07$~mm$^{-1}$. Fitting results show that $d = 5$~$\mu$m structures don't exhibit coupling coefficient damping but preserve a highly anisotropic behaviour.
	}
	\label{fig:dc_calibration}
\end{figure}

We fabricated 5 arrays of PDCs with different distance $d$ between the waveguides inside the interaction region. Each array consists of directional couplers with identical $d$ and varied interaction length $L$. The polarization properties of the fabricated PDC structures were studied with the simple transmission measurement setup. Linearly polarized (H or V) laser light at 808 nm was sent to one of the input ports of the directional coupler. The light from the output ports of the structures was collimated and sent to the power meter positioned $\approx$ 1.5 m apart from the collimating lens, thus ensuring the measured light power corresponds to the guided mode only. The results are presented in Fig.\ref{fig:dc_calibration}. We observe weak polarization sensitivity in the PDC structures with $d = 6$ $\mu$m and $d = 7$ $\mu$m. 

In order to minimize the footprint of the polarizing coupler it is desirable to reduce $L_{pdc}$ as much as possible, i.e. to increase $\left|C_V-C_H\right|$. For almost symmetric waveguides the only way to introduce stronger anisotropy is to reduce the distance between the waveguides. However, placing the waveguides close to each over comes at a price: as may be clearly seen from Figs.~\ref{fig:3mkm_pdc},~\ref{fig:5mkm_pdc} the birefringence indeed increases for closer waveguides, but the oscillations of power between the waveguides are damped. Moreover, this effect is becoming more pronounced for the couplers with smaller distance between the waveguides, as evidenced by Fig.~\ref{fig:3mkm_pdc}. We attribute this phenomenon to the manifestation of the irregularities appearing in the process of writing a waveguide in close proximity to another one. 

Let us introduce a theoretical model that explains the observed behaviour. In the following we will consider a single polarization mode and omit the polarization indices. Evolution of the field amplitudes $a_1(z)$ and $a_2(z)$ in the two arms of the coupler may be described by the coupled modes equations \cite{Snyder_Love}:
\begin{equation}
\label{eq:coupled_modes}
	\left\{
		\begin{split}
			&\frac{da_1}{dz}=-i\beta_1(z)a_1-iCa_2,\\
			&\frac{da_2}{dz}=-iCa_1-i\beta_2(z)a_2,
		\end{split}
	\right.
\end{equation}
where $\beta_{j}(z)$ is the wavenumber dependent on the propagation coordinate due to the irregular modification of the mode’s effective refractive indices ($j=1,2$). Equations (\ref{eq:coupled_modes}) do not take loss in the waveguides into account, since it is trivial to do under the plausible assumption of equal loss. Thus, one can show that the total optical power is conserved: $\left|a_1(z)\right|^2+\left|a_2(z)\right|^2=\left|a_1(0)\right|^2+\left|a_2(0)\right|^2$ and the evolution, described by (\ref{eq:coupled_modes}) is unitary.

One can split the wavenumber into to parts: $\beta_j(z)=\beta_{j0}+\delta\beta_j(z)$, with $\beta_{j0}$ being the constant part and $\delta\beta_j(z)$ -- the variable part, which is fluctuating irregularly. The precise form of the irregular part is unknown, but we assume, that the characteristic length $l_0$ over which $\delta\beta_j(z)$ changes significantly is much smaller, than the period of the amplitude oscillations induced by the evanescent coupling $L_0=\pi/2C$. Under this assumption, equations (\ref{eq:coupled_modes}) can be regarded as stochastic ones with $\delta\beta_j(z)$ playing the role of random fluctuations. Since $l_0\ll L_0$ the fluctuation properties are defined by the correlation functions: $\langle\delta\beta_j(z)\rangle=0$, $\langle\delta\beta_j(z)\delta\beta_k(z')\rangle=2D_{jk}\delta(z-z')$, where $\delta(x)$ is the Dirac delta function and parameters $D_{jk}$ quantify the fluctuations intensity.

We take the coupling coefficient $C$ to be constant, while accounting for possible variations in wavenumbers, which requires some justification. Indeed, the coefficient is calculated using (\ref{eq:coupling_C}) where both the refractive index contrast and the eigenmode profiles are influenced by fluctuations of the refractive index. Therefore small variations of the refractive index contrast $\delta(\Delta n)\ll \Delta n$ imply small deviations in the coupling coefficient $\delta C\ll C$. Similarly, the same variations $\delta(\Delta n)$ yield a change in the wavenumber $\delta\beta_j=k\delta n_{eff}\ll \beta_j=kn_{eff}$, which however, may be comparable to the values of $C$ due to the typically large values of $\beta_j$, and thus may significantly influence the dynamics. 

To calculate the statistical averages $p_j(z)=\left\langle | a_j(z)|^2 \right\rangle$, it is convenient to first derive the equations for the quadratic quantities – the power difference $\Delta(z)=\left|a_1(z)\right|^2 - \left|a_2(z)\right|^2$ and the complex-valued product $\sigma(z)=a_1(z)a_2^*(z)$, which can be done straightforwardly using Eqs.~\ref{eq:coupled_modes}. Statistical averaging may be performed analytically under the assumption of noise gaussianity: $\left\langle\exp(i\Phi(z)\right\rangle=\exp\left(-\frac{1}{2}\langle\Phi^2(z)\rangle\right)$. We arrive to the following set of equations for three real-valued functions $\Delta(z)$, $\sigma'=\mathrm{Re}\sigma$ and $\sigma''=\mathrm{Im}\sigma$:
\begin{equation}
\label{eq:Bloch}
	\left\{
	\begin{split}
		&\frac{d\langle\sigma\rangle}{dz}=iC e^{-Dz}e^{-\beta z}\langle\Delta\rangle,\\
		&\frac{d\langle\Delta\rangle}{dz}=-4C e^{-Dz}\left( \langle\sigma'\rangle \sin(\beta z) + \langle\sigma''\rangle \cos(\beta z) \right),
	\end{split}
	\right.
\end{equation}
where $\beta=\beta_{20}-\beta_{10}$ and $D=D_{11}+D_{22}-D_{12}$. Let us note, that $D$ is the only parameter describing the fluctuations which enters the equations and may be obtained from the fit of the experimental data. Finally, one can obtain a single equation for the power difference from (\ref{eq:Bloch}):
\begin{equation}
\label{eq:power_difference}
	\frac{d^3\Delta}{dz^3}+2D\frac{d^2\Delta}{dz^2}+\left( 4C(z)^2+D^2+\beta^2 \right) \frac{d\Delta}{dz} -4C(z)^2D\Delta = 0,
\end{equation}
where $C(z)=C\exp(-Dz)$ is the coupling coefficient, damped due to the propagation constant fluctuations. These equations should be solved with the following initial conditions: 
\begin{equation*}
	\begin{split}
		&\Delta(0)=p_{10}-p_{20}=(2\xi-1)p_0,\\ &\Delta'(0)=-4C\sqrt{\xi(\xi-1)}p_0\sin\Delta\varphi, \\
		&\Delta''(z)=-4Cp_0\left( \sqrt{\xi(1-\xi)}\left[D\sin\Delta\varphi+\beta\cos\Delta\varphi\right]\right. \\
		&\qquad\quad\,\,\,\left.+C(2\xi-1)  \right).
	\end{split}
\end{equation*}
Here $\xi=p_{10}/p_0$ is the fraction of total power at the input of the first waveguide and $\Delta\varphi=\arg(a_{10}a_{20}^*)$ is the phase difference between the input amplitudes.

\begin{figure}[htb]
	\subfloat[]{
		\centering
		\includegraphics[width=0.49\linewidth]{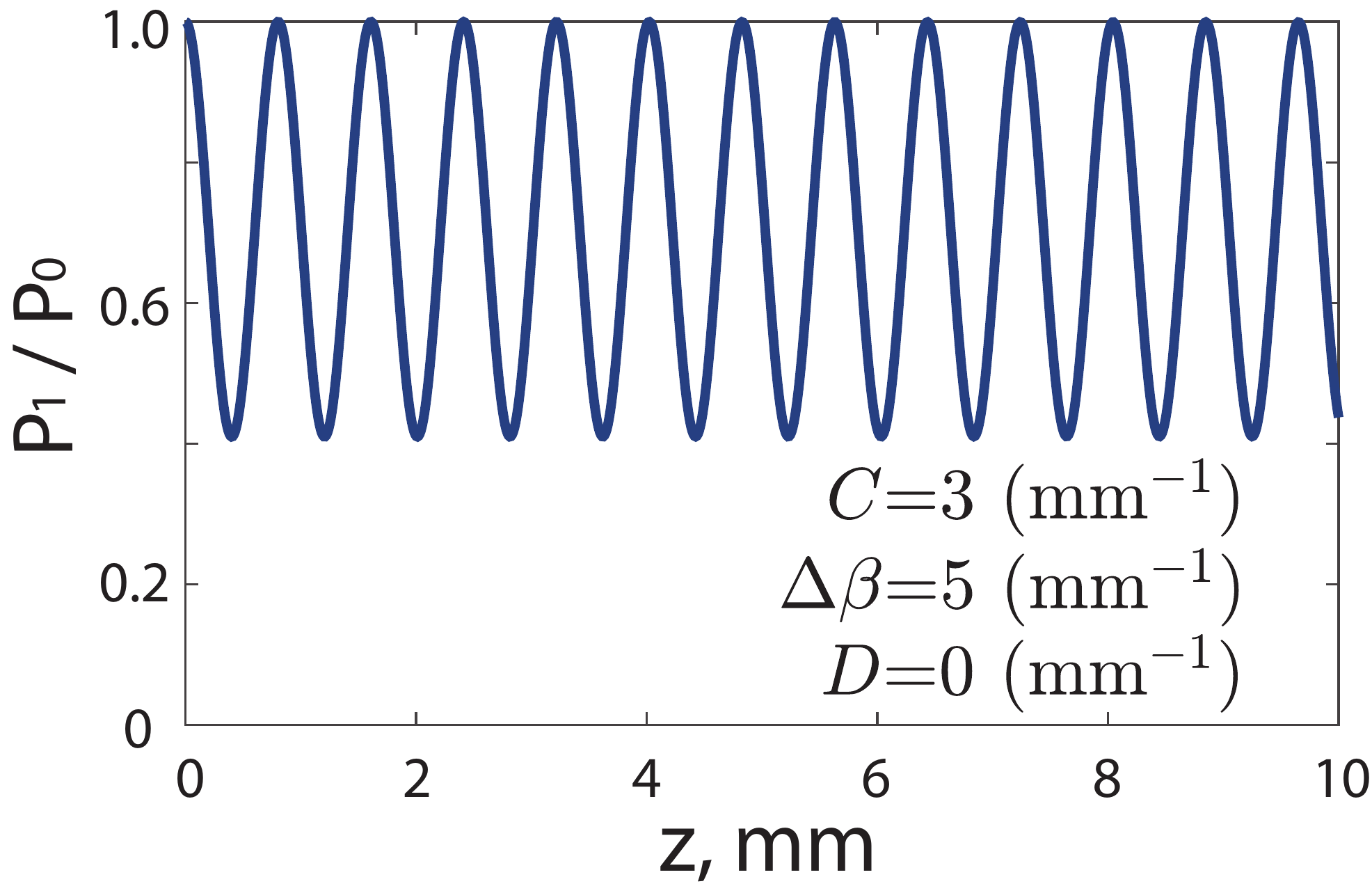}
		\label{fig:non_zero_beta}
	}
	\subfloat[]{
		\centering
		\includegraphics[width=0.49\linewidth]{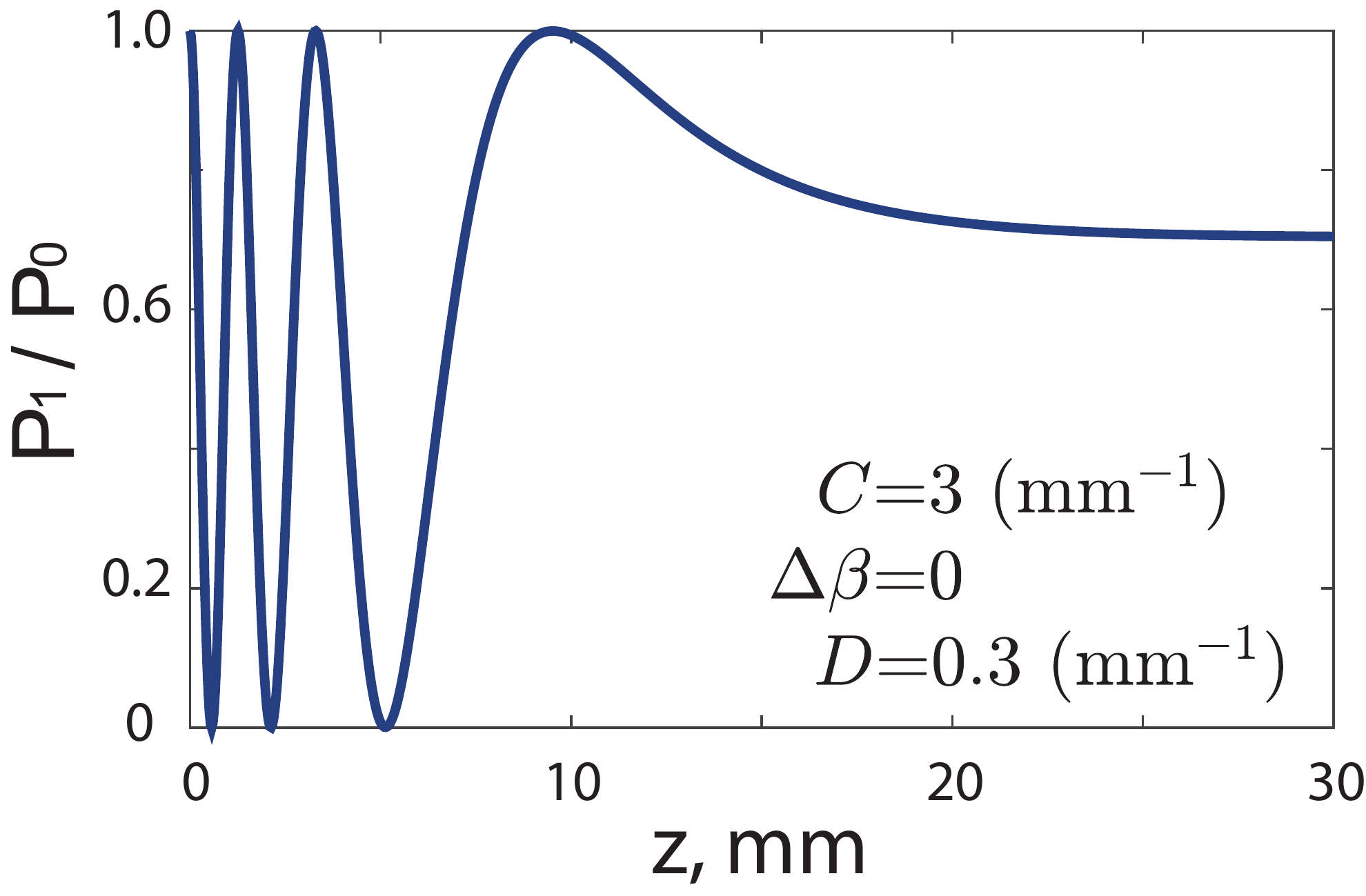}
		\label{fig:non_zero_noise}
	}\\
	\subfloat[]{
		\centering
		\includegraphics[width=0.49\linewidth]{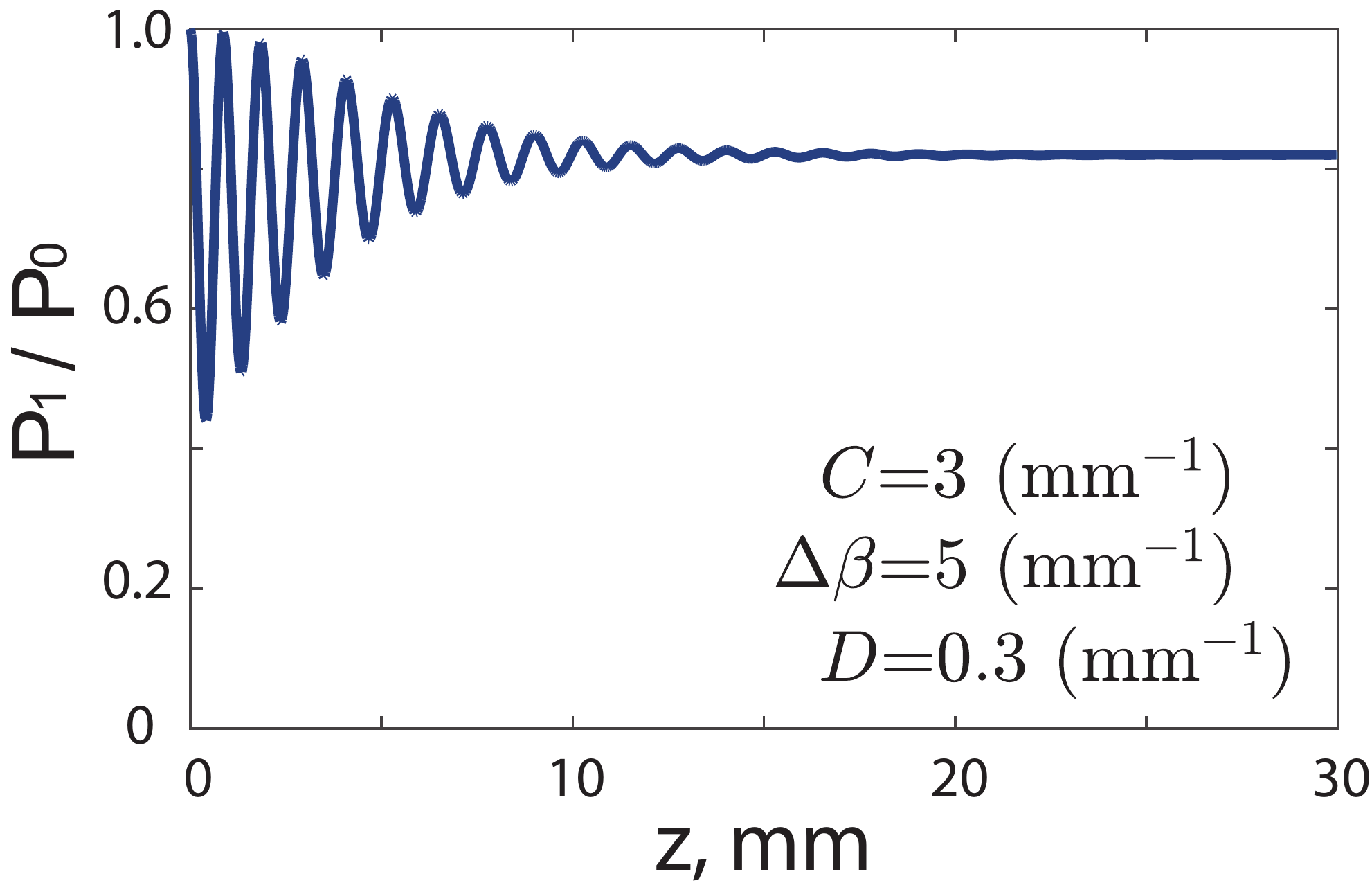}
		\label{fig:non_zero_both}
	}
	\subfloat[]{
		\centering
		\includegraphics[width=0.49\linewidth]{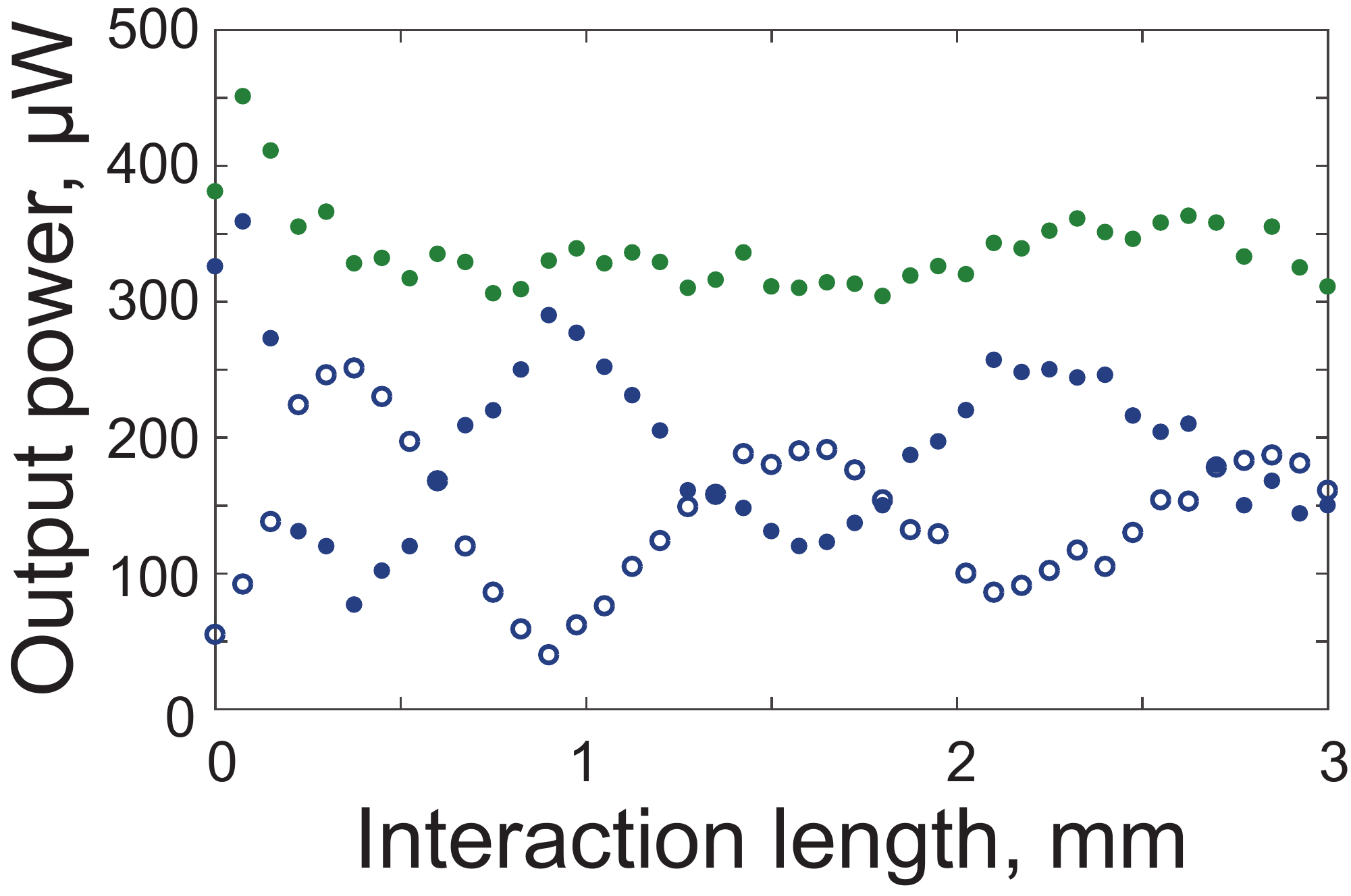}
		\label{fig:raw_3mkm}
	}
	\caption{Dependence of the fraction of power in the first waveguide on the interaction length, obtained by numerically solving (\ref{eq:power_difference}). \ref{fig:non_zero_beta}--\ref{fig:non_zero_both} illustrate the behaviour of the solutions depending on various choices of parameters. \ref{fig:raw_3mkm} shows the measured power in both output arms of the directional coupler (blue dots and circles) and their sum (green dots) to illustrate the conservation of energy in the guided modes.}
	\label{fig:example_theory_figures}
\end{figure}

We were unable to find an analytical solution for Eq.~\ref{eq:power_difference} in the general case. The influence of each of the parameters on the power dynamics in the waveguides is illustrated in Fig.~\ref{fig:example_theory_figures}. Non-zero detuning $\beta$ with no fluctuations ($D=0$) leads to an incomplete power transfer between the waveguides (Fig.~\ref{fig:non_zero_beta}). In the opposite case of $\beta=0$ and $D\neq0$ (Fig.~\ref{fig:non_zero_noise}) random fluctuations prohibit the power transfer, causing the exponential decay of the effective coupling coefficient $C(z)$. When both coefficients are non-zero, as shown in Fig.~\ref{fig:non_zero_both}, the behaviour of the power distribution resembles the one observed in the experiment. We used the numerical solution of (\ref{eq:power_difference}) to fit the experimental data in Fig.~\ref{fig:dc_calibration}. The parameters, corresponding to the best fit, are given in the figure caption. Let us note, that the observed behaviour can not be described by radiative losses in the coupled waveguides, since the total power in both waveguides is conserved, as shown in Fig.~\ref{fig:raw_3mkm}. 

\begin{figure}[!htb]
	\centering
	\includegraphics[width=\linewidth]{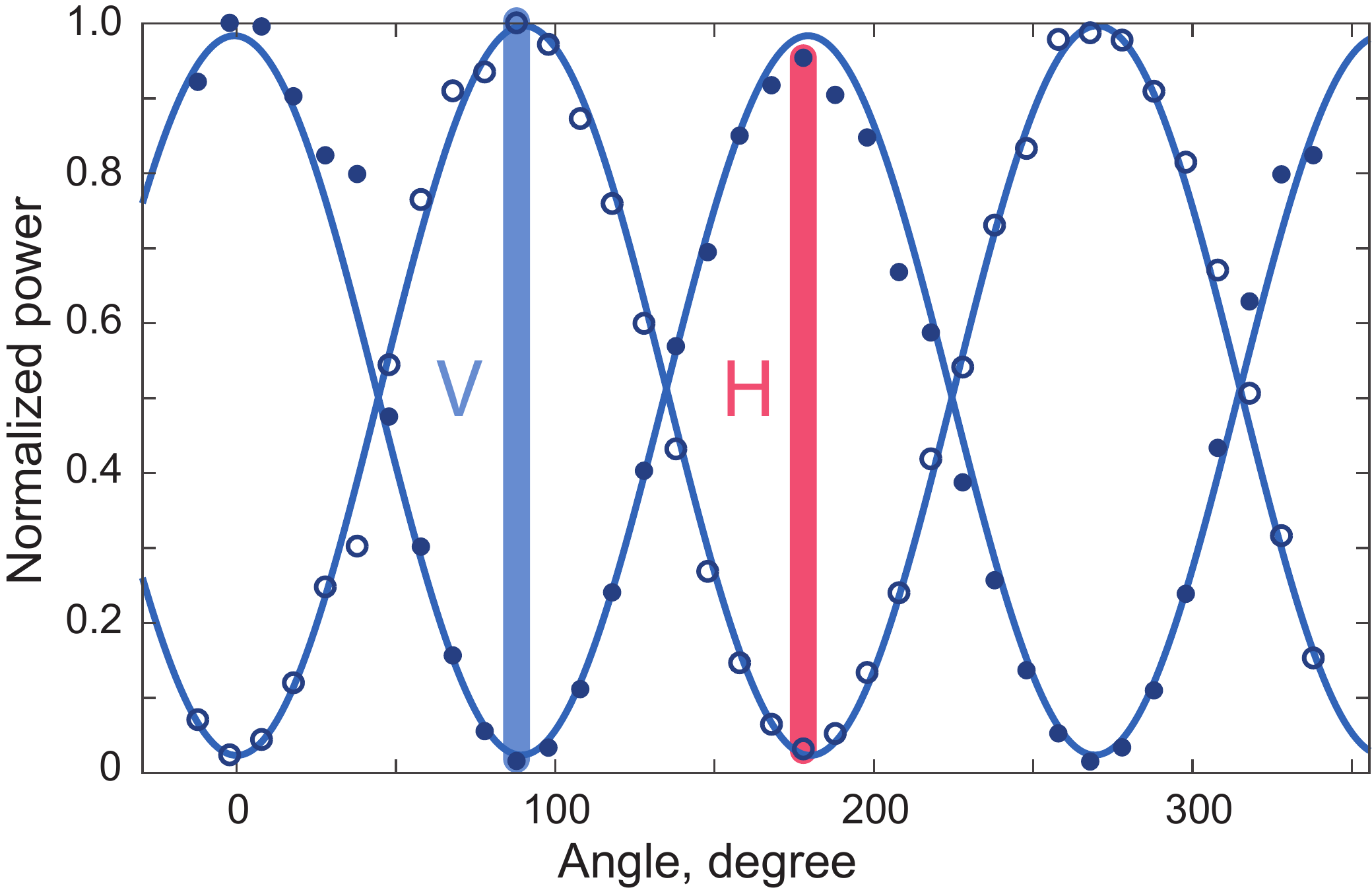}
	\caption{Polarization extinction measurement for a completely polarizing integrated DC. Full circles represent the relative intensity in the reflected arm of the PDC, open circles -- in the transmitted arm. All data points are normalized to the maximal output intensity. Solid lines represent the Malus' law curve with $R_{H}$, $R_{V}$ or $T_{H}$, $T_{V}$ as the fitting coefficients, correspondingly.}
	\label{fig:malus}
\end{figure}
The quality of the fabricated PDC ($d = 5$ $\mu$m, $L = 3.7$ mm) was established in the extinction measurement. We set the polarization state of the input laser beam by the Glan-Thompson polarizer and measured the intensity in each of the output ports of the structure. The dependencies obtained are shown in Fig.~\ref{fig:malus}. The extinction ratios of the PDC may be inferred by fitting the obtained dependency with the Malus' law equation accounting for non-ideal $R_{H,V}$, and $T_{H,V}$ and takes values of 20 dB for the vertical and 16 dB for the horizontal input polarization state. Repeatable fabrication of high extinction integrated PDC devices is still a challenging task due to intrinsic random defects occurring during the writing process. That leads to fluctuations in the parameters of the couplers, clearly observed for longer interaction lengths in Fig.~\ref{fig:5mkm_pdc}. However, to the best of our knowledge, the device reported in this work is the shortest polarizing directional coupler fabricated with femtosecond laser writing technology.

We have experimentally investigated the anisotropically coupled FSLW waveguides in fused silica with spacing between the waveguides as small as 3~$\mu$m. Strong anisotropic coupling is observed for distances between the waveguides below 5~$\mu$m, leading to the reduced interaction length, required to realize a polarizing directional coupler. The demonstrated coupler has the measured extinction ratios of 16 dB and 20 dB for the horizontal and vertical polarizations, respectively, comparable to the state of the art FSLW integrated devices, but with an order of magnitude lower interaction length of $3.7$~mm. The reduced footprint of the polarizing elements manufactured with our technique, together with low intrinsic birefringence of the interconnecting waveguides, makes our approach favourable for building complex polarization-sensitive integrated circuits. 

Stronger anisotropic coupling for closely spaced waveguides is accompanied by some detrimental effects, manifesting themselves in the reduction of visibility of power oscillations for the coupled modes. We have observed the effect experimentally and provided a model, explaining the basic features of the observed behaviour. Although the model qualitatively reproduces the observed effects, the underlying assumptions have to be tested in independent experiments, which we leave for future work. The observed effects of suppression of the power transfer for longer directional couplers with small spacing may be used to design couplers with unusual properties, such as the spectral dependence of the splitting ratio, these opportunities will be investigated elsewhere. 

\textbf{Funding.}
Russian Science Foundation project 16-12-00017, theoretical work of M.Yu.~Saygin is supported by RSF project 17-72-10255.

\bibliography{ipbs_bibliography}


\end{document}